\documentclass[12pt]{article}

\begin{document}

\sloppy
\begin{flushright}{SIT-HEP/TM-5}
\end{flushright}
\vskip 1.5 truecm
\centerline{\large{\bf Affleck-Dine baryogenesis after}}
\centerline{\large{\bf thermal brane inflation }}
\vskip .75 truecm
\centerline{\bf Tomohiro Matsuda
\footnote{matsuda@sit.ac.jp}}
\vskip .4 truecm
\centerline {\it Laboratory of Physics, Saitama Institute of
 Technology,}
\centerline {\it Fusaiji, Okabe-machi, Saitama 369-0293, 
Japan}
\vskip 1. truecm
\makeatletter
\@addtoreset{equation}{section}
\def\theequation{\thesection.\arabic{equation}}
\makeatother
\vskip 1. truecm

\begin{abstract}
\hspace*{\parindent}
  We propose a new scenario of Affleck-dine baryogenesis in the context
 of theories with large extra dimensions. In this paper we consider
 baryogenesis after thermal brane inflation and show how our mechanism
 works. We specifically consider models in which supersymmetry is broken
 at the distant brane. 
\end{abstract}

\newpage
\section{Introduction}
\hspace*{\parindent}
The production of net baryon asymmetry requires baryon number violating
interactions, C and CP violation and a departure from thermal 
equilibrium\cite{sakharov}.
The first two of these ingredients are naturally contained in GUTs or
other string-motivated scenarios, and the 
third can be realized in an expanding universe where it is not uncommon that
interactions come in and out of equilibrium, producing the stable
heavy particles or cosmological defects.
In the original and simplest model of baryogenesis\cite{original},
a heavy GUT gauge or Higgs boson decays out of equilibrium producing
a net baryon asymmetry.

Another mechanism for generating the cosmological baryon asymmetry
in supersymmetric theories is proposed by Affleck and 
Dine\cite{AD} who utilized the decay of the scalar condensate along the
flat direction. 
This mechanism is a natural product of supersymmetry, which contains
many flat directions that break $U(1)_{B}$.
The scalar potential along this direction vanishes identically when 
supersymmetry breaking is not induced.
Supersymmetry breaking lifts this degeneracy,
\begin{equation}
V\simeq m^{2}_{soft}|\phi|^{2}
\end{equation}
where $m^{2}_{soft}$ is the supersymmetry-breaking scale and
$\phi$ is the direction in the field space corresponding to
the flat direction.
For large initial value of $\phi$, a large baryon number asymmetry
may be generated if the condensate of the field breaks $U(1)_{B}$.
The mechanism also requires the presence of baryon number violating 
operators that may appear through higher dimensional A-terms.
The decay of these condensates through such an operator can lead to a 
net baryon asymmetry.
In the most naive consideration the baryon asymmetry is computed by 
tracking the evolution of the sfermion condensate in the flat direction
of the supersymmetric standard model.
Considering a toy model with the potential
\begin{equation}
V(\phi,\phi^{\dagger})=m^{2}_{soft}|\phi|^{2}+\frac{1}{4}[\lambda
\phi^{4}+h.c.],
\end{equation}
the equation of motion becomes
\begin{equation}
\label{toy}
\ddot{\phi} + 3H\dot{\phi} = - m_{soft}^2 \phi 
+ \lambda (\phi^{\dagger})^{3}.
\end{equation}
The baryon (or lepton) number density is given by;
\begin{equation}
n_{B}=q_{B}\left(\dot{\phi}^{\dagger}\phi-\phi^{\dagger}\dot{\phi}\right),
\end{equation}
where $q_{B}$ is the baryon (or lepton) charge carried by the field.
Now one can write down the equation for the baryon number density
\begin{equation}
\dot{n_{B}}+3Hn_{B}=2q_{B}Im\left[\lambda (\phi^{\dagger})^{4} 
\right].
\end{equation}
Integrating this equation, one can obtain the baryon (or lepton)
number produced by the Affleck-Dine oscillation.
For a large initial amplitude, the produced baryon number is estimated as
$n_{B} \simeq \frac{4q_{B}|\lambda|}{9H} |\phi_{ini}|^4 \delta_{eff}$,
where $\delta_{eff}$ is the effective CP violation phase of the initial
condensate.
This crude estimation suggests that by generating some angular motion 
one can generate a net baryon density.

In the conventional scenario of Affleck-Dine baryogenesis, one should
assume large $H>m_{3/2}$ before the time of Affleck-Dine baryogenesis 
so that the flat directions are destabilized to obtain the large initial
amplitude of baryon-charged directions.

Although it seems plausible that Affleck-Dine baryogenesis generates
the baryon asymmetry of the Universe, there are some difficulties in the 
naive scenario.
The formation of Q-ball\cite{kusenko} is perhaps the most serious obstacle
that puts a serious constraint on the baryon number density at the
time of Q-ball formation.
Q-balls are formed due to the spatial instability of the Affleck-Dine
field, and have been shown by numerical calculations that they absorb
almost all the baryonic charges in the Universe when they
form\cite{Kasuya, Enq}.
This means that the baryon asymmetry of the Universe in the later period
must be provided by decaying Q-balls.
In general, the stability of Q-balls are determined by their charge
that are inevitably fixed by Affleck-Dine mechanism itself.
The reason is that the formation of Q-balls occurs almost immediately,
which makes it hard to expect any additional diluting mechanism before
Q-ball formation.
The point is that in general Affleck-Dine baryogenesis the initial
baryon number density becomes so huge that the produced Q-balls become
 stable.
The stable Q-balls that produce the present baryon asymmetry of the
Universe by their decay are dangerous, because such Q-balls can also produce
dangerous relics at the same time when they decay to produce the baryons.
The decay temperature of the associated huge Q-balls becomes in general 
much lower than the 
freeze-out temperature of the dangerous lightest supersymmetric
particle, which causes serious constraint.

Another obstacle is the problem of the early oscillation caused by the
thermalization\cite{dineT}.
When the fields that couple to the Affleck-Dine field are thermalized,
they induce the thermal mass to the Affleck-Dine field.
The early oscillation starts when the thermal mass term exceeds the
destabilizing mass.
The serious constraint appears because the destabilizing mass, which
is about the same order of the Hubble parameter, is in general 
much smaller than the temperature of the plasma.

However, in our model these difficulties do not appear
since the mechanism of the destabilization of the Affleck-Dine field is
not a consequence of large Hubble parameter.
The size of the Q-ball is naturally suppressed, since our mechanism does not 
produce huge baryon number density after Affleck-Dine baryogenesis.

What we will consider in this paper is a mechanism in which
Affleck-Dine mechanism is realized after thermal brane 
inflation\cite{weak_brane}.
Before discussing the baryogenesis with extra dimensions, we must first
specify the scenario of the early Universe to a certain extent.
In this paper we consider Affleck-Dine baryogenesis after thermal
brane inflation.\footnote{
In theories with extra dimensions there are two possible choices for the
Affleck-Dine field.
It could either be a brane field localized on a brane or a bulk field.
In ref.\cite{AD_extra}, it is discussed that naive realization of
Affleck-Dine 
mechanism with a brane field cannot produce sufficient baryon number.}
We show how our mechanism works in models with supersymmetry breaking at
the distant brane.
Here we consider two different cases for supersymmetry breaking.
In one case we assume an alternative source of supersymmetry breaking
on the distant brane, and in the other case we deal with the realistic
bulk field mediation of supersymmetry breaking. 
It is often the case that the brane distance is used to control the
direct contact terms that produce unwanted soft terms preventing the
FCNC bound.
Our mechanism is expected to work in these models if the relevant brane
distance is reduced during a period after inflation.
Besides the thermal inflation model that we have considered in this paper, 
there are many models in which the temporally reduced extra dimension is
used to prevent difficulties related to the large extra
dimensions\cite{other}.

\section{Thermal brane inflation}
\hspace*{\parindent}
In this section we make a brief review of thermal brane
inflation proposed by Dvali\cite{weak_brane}.
The following conditions are required so that the mechanism functions.

1) \,Exchange of the bulk modes such as graviton, dilaton or RR
fields governs the brane interaction at the large distance.

2)\, In the case when branes initially come close, bulk modes are in
equilibrium and their contribution to the free energy
can create a positive $T^{2}$ mass term for $\phi$ to stabilize the
branes on top of each other until the Universe cools down to a certain
critical temperature $T_{c}\sim m_{s}$.\footnote{
Authours of ref.\cite{weak_brane}
 considered open string modes stretched between the different
branes.
If the branes are on top of each other, these string modes that get
mass when the brane distance grows are in equilibrium and their
contribution to the free energy creates a positive $T^2$ mass term
so that the resulting curvature becomes positive. }
Here $m_s$ represents the negative curvature of $\phi$ at the origin
determined by the supersymmetry breaking, and $\phi$ is the moduli field 
for the brane distance.

The resulting scenario of thermal inflation is straightforward.
Assuming that there was a period of an early inflation with a reheat
temperature $T_{R}\sim M$, and at the end of inflation some of the
repelling branes sit on top of each other stabilized by the thermal
effects, one can obtain the number of e-foldings 
\begin{equation}
N_e=ln(\frac{T_{R}}{T_{c}}).
\end{equation}
Taking $T_{R}\sim 10 TeV$ and $T_{c}\sim 10^{3}- 10$ MeV, one finds 
$N_{e}\sim 10-15$, which is consistent with the original thermal
inflation\cite{stewart} and is enough to get rid of unwanted 
relics.\footnote{
In this paper we also consider situations where the reheating temperature
after the first inflation is as high as $T_{R} \sim 10^{10}GeV$,
and the critical temperature $T_{c} \sim 10^{2}GeV$.
Unlike the original model of thermal brane inflation, large extra 
dimensions are not specially supposed in this paper.
Since we are taking interest in whether our mechanism of baryogenesis 
works, 
we also deal with the case where the thermal brane inflation itself is not 
a necessary ingredient to solve the cosmological problems.
In such cases, what we should concern is whether there can be a short
period of thermal brane inflation that enables our mechanism of Affleck-Dine
baryogenesis to work.}

\section{Affleck-Dine baryogenesis after thermal brane inflation}
\hspace*{\parindent}
In this section we show how to realize Affleck-Dine baryogenesis after
thermal brane inflation.
Our model requires the mechanism of supersymmetry breaking 
at the distant brane.
To proceed, we should first discuss the mechanism of supersymmetry
breaking.
In our model the negative soft term is not a simple consequence of the
large Hubble parameter, but rather related to the distance between the
matter brane and the supersymmetry-breaking brane.
We should also discuss the origin of the baryon number violating A-terms,
which plays the crucial role in Affleck-Dine baryogenesis.
Because of the constraint from proton stability, additional mechanism
for suppressing dangerous higher dimensional A-terms is always required when
the fundamental scale is much lower than the Planck scale.

In the oldest version of supergravity mediation, it is assumed that all
higher-dimension operators that directly connect the fields in the
hidden sector with the ones in the observable sector
 are present but suppressed only by powers
of $1/M_{4}$, where $M_4$ denotes the Planck mass in four dimensions. 
In this case the required soft supersymmetry-breaking term is given by
the higher-dimension terms of the form:
\begin{equation}
\label{soft}
L_{soft} \sim \int d^{4}\theta\,
\frac{1}{M_{4}^{2}} X^\dagger X Q^\dagger Q
\end{equation}
Here $X$ is a chiral superfield in the hidden sector whose $F$ component
$F_X$ breaks supersymmetry.
$Q$ is a matter field in the visible sector.
Higher dimensional operators in the superpotential $W_A \sim
\frac{1}{M_p^{n+3}}\Phi^{n+3}$ produce the A-terms and determines the
phase of the AD direction at large $<\Phi>$.  
\begin{equation}
\label{A-term}
L_{A} \sim 
\int d^{4}\theta\,\left(
\frac{1}{M_4^{n+3}} X^\dagger X\Phi^{n+3} + h.c. \right)
+\int d^{2}\theta\,\left(
\frac{1}{M_4^{n+1}} X\Phi^{n+3} + h.c. \right)
\end{equation}
where $n \ge 1$ and $\Phi$ represents the flat direction.

Contact terms of the similar form appear in the models of
extra dimensions, where $M_4$ is replaced by the fundamental scale $M$
that is much lower than $M_4$.
On the other hand, the contact terms connecting the fields in the hidden
and the observable branes are suppressed because they are localized
along the extra dimension.
In these cases the supersymmetry breaking is mediated by bulk fields
such as scalar fields\cite{anomaly_mediation,radion_mediation} or
fermions\cite{gaugino_mediation} where the scale of the supersymmetry
breaking in the hidden brane can be as large as the fundamental scale of
the higher-dimensional theory, while the direct soft terms for the standard
model sfermions are suppressed.

\underline{Simplest toy model}

For the simplest toy model we consider an example where the fundamental
scale $M$ is as low as $10TeV$ and the realistic supersymmetry breaking
is realized within the matter brane without specific fine-tuning.
In addition to these simplest settings, we also include a distant brane
where the supersymmetry is maximally broken by an auxiliary component
of a localized field $|F_X|^{1/2} \sim M$.
In such a case the effect of $F_X$ on the matter brane is expected to be
exponentially suppressed because they are localized at the distant brane.
The soft terms are given by:
\begin{equation}
V(\phi_{AD}) \sim \left[m_{soft}^2
+c \left(\frac{|F_X|}{M}\right)^2 e^{-Mr_{susy}}\right]
|\, \phi_{AD}|^2.
\end{equation}
Here $\phi_{AD}$ is the flat direction of Affleck-Dine mechanism,
and $r_{susy}$ is the distance between the matter brane and the hidden
supersymmetry-breaking brane on which $F_X$ is localized.
$m_{soft}$ denotes the supersymmetry breaking induced on the matter brane,
which is assumed to be a constant.
When two branes sit on their true positions, the second term is
negligible.
On the other hand, when the hidden brane stays on top of the matter brane
during thermal brane inflation, then the supersymmetry breaking of order 
$F_X /M$ is induced  on the matter brane by the direct contact terms.
Assuming that the effective soft mass appears with the negative 
sign (i.e. $c<0$),
the flat direction $\phi_{AD}$ is destabilized during thermal inflation
if $m_{soft}<|F_X|/M$.
At the same time A-terms are modified to generate the required
misalignment of the phase.
Here we assume that the A-term is effectively given by using the four
dimensional Planck mass,
\begin{equation}
V_A \simeq 
\left(\frac{a_0 m_{soft}}{M_p}+
\frac{a_1 |F_X| e^{-Mr_{susy}}/M}{M_p}\right)
\phi_{AD}^4
\end{equation}
where $a_0$ and $a_1$ are constants of O(1).
The situation here is very similar to the original Affleck-Dine
baryogenesis.
The sole difference is that the supersymmetry is not induced by the 
Hubble parameter, but is induced by the brane distance.
The resultant baryon to entropy ratio is:\footnote{
See ref.\cite{AD_dine} for more detail.}
\begin{equation}
\frac{n_{B}}{s}\sim \frac{T_{R2}}{M_p H_o \rho_I}
|a m_{soft} (\phi_{AD}^{i})^4 |\delta_{eff}
\end{equation}
where $T_{R2}$ is the reheating temperature after thermal brane
inflation, and $\phi_{AD}^i$ is the initial amplitude of
$\phi_{AD}$.
$H_o$ denotes the Hubble parameter when the AD oscillation starts,
which can be taken to be $H_o \leq H_I=M^2/M_p$.
It is naturally assumed that the initial amplitude is  
$\phi_{AD}^{ini}\sim M$, and the inflaton density is still $\rho_{I}
\sim M^4$ at the beginning of the oscillation.
Then we obtain:
\begin{equation}
\frac{n_{B}}{s}\sim 10^{-10} \left(\frac{T_{R2}}{10MeV}\right)
\left(\frac{10^{-8} GeV}{H_o}\right)
\end{equation}
which is the most naive result, but is enough to explain the origin of
the baryon asymmetry of the present Universe.\footnote{
Modifications of parameters are allowed, but in general they are 
strongly model dependent.
The magnitude of the A-term can be modified at the time of AD
oscillation, which we shall discuss in the next paragraph.}

\underline{Realistic mediation of supersymmetry breaking}

Here we consider gaugino mediation as a more realistic example of such 
``Hidden'' supersymmetry breaking\footnote{
Here we have assumed that the
gaugino can propagate only one
extra dimension that is about $10-10^{2}$ times larger than
$M^{-1}$\cite{gaugino_mediation}. }.
In this case the MSSM scalar mass squareds derived from five-dimensional
Feynman diagrams are suppressed relative to the gaugino masses by at
least a loop factor when the brane distance is larger than $M^{-1}$.

Even in the limit of small brane distance, it does not exceed the masses
generated from the four-dimensional renormalization-group
evolution between the compactification scale and the weak 
scale\footnote{Details of the calculations are given in ref.\cite{Peskin}}.
This conclusion is generic and also holds for the other soft parameters
such as A-terms.

Besides the contributions from five-dimensional Feynman diagrams
of gauginos propagating through extra dimension, there are direct
contact terms that can destabilize the flat direction during thermal
brane inflation.

Here we consider two sources of supersymmetry breaking,
the four-dimensional effect and the direct contact term.
Assuming that the soft terms of the relevant flat direction is 
produced by these two sources, 
it takes the following form\cite{gauge_mediation}:
\begin{equation}
\label{gauge_med_soft}
V(\phi_{AD}) \sim \left[c_1 \left(\frac{g^2_4}{(4\pi)^2}\right)^2 
\left(\frac{|F_X|}{M}\right)^2 
+c_3 \left(\frac{|F_X|}{M}\right)^2 e^{-Mr_{susy}}\right]|\, \phi_{AD}|^2
\end{equation}
for small $\phi_{AD}$ and
\begin{equation}
V(\phi_{AD}) \sim c_2 \left(\frac{g^2_4}{(4\pi)^2}\right)^2 (|F_X|)^2 
\left(ln\frac{|\phi_{AD}|^2}{M^2}\right)^2
+c_3 \left(\frac{|F_X|}{M}\right)^2 e^{-Mr_{susy}}\,|\phi_{AD}|^2
\end{equation}
for large $\phi_{AD}$.
Here $\phi_{AD}$ is the flat direction of Affleck-Dine mechanism,
and $M$ is the fundamental scale.
$r_{susy}$ is the distance between the supersymmetry-breaking brane and 
the matter brane.

If the supersymmetry-breaking hidden brane stays on top of our brane
during thermal brane inflation, the supersymmetry breaking
on our brane during this period is naturally the order of $O(|F_{X}|/M)$ because
the $e^{-Mr_{susy}}$ factor in the direct contact term is $O(1)$.
Relevant soft mass is then given by:
\begin{equation}
m^2(\phi_{AD}) \sim c_{1}\left(\frac{g^2}{(4\pi)^2}\right)^2 
\left(\frac{|F_X|}{M}\right)^2
+c_{3}\left(\frac{|F_{X}|}{M}\right)^{2} e^{-Mr_{susy}}
\end{equation}
for $\phi_{AD}<M$.
It is obvious that the  source of supersymmetry breaking during
thermal brane inflation is in general different from the one at the true
vacuum.
At this time the flat directions on the observable brane are lifted or
destabilized by the supersymmetry breaking induced by the direct contact
terms, which will soon disappear as soon as the brane distance grows.
Then the situation is similar to the conventional Affleck-Dine
baryogenesis, where the destabilization is induced by the alternative
supersymmetry breaking induced by the inflaton.
Assuming that the direct supersymmetry breaking destabilizes the flat
direction with the negative soft mass, which corresponds to taking 
the constant $c_3<0$, the potential of the
Affleck-Dine flat direction at the end of thermal inflation is
given by
\begin{equation}
V_{soft}(\phi_{AD})\sim -|c_3|\left(\frac{|F_X|}{M}\right)^2|
\phi_{AD}|^{2}.
\end{equation}
This negative soft mass disappears soon after the end of thermal brane 
inflation, as the brane distance $r_{susy}$ grows.

The direct contact terms decreases exponentially, while terms produced
by the four-dimensional effect are not modified by $r_{susy}$, because
the supersymmetry-breaking gaugino mass is determined by the size of the
relevant extra dimension that is assumed to be a constant during
thermal brane inflation.
Then there should be an oscillation of the Affleck-Dine field 
that starts at $r_{susy}\sim M^{-1}$.

We should also consider another important ingredient of Affleck-Dine
baryogenesis, the evolution of the A-term.
To discuss the magnitude of A-terms during thermal brane inflation,
we must first discuss a concrete model for suppressing the baryon number
violating interactions.
The most naive idea is to assume that the baryon number is maximally
broken on the hidden brane and its effect on our brane is exponentially
suppressed by $e^{-r_B}$  where $r_B$ is the distance between the 
hidden brane
and the matter brane\cite{brane_baryo,e_n_p_decay,shine}.
A popular mechanism for explaining the smallness of the observed
Yukawa couplings or baryon number violating operators is to expect
higher dimension operators of the generic form:
\begin{equation}
{\cal O} \sim \lambda \left(\frac{\chi}{M}\right)^k {\cal O}_{MSSM}
\end{equation}
with $\lambda\sim O(1)$.
If $\epsilon \sim \frac{\chi}{M}$ is small, the small couplings in these
operators are understood as the small parameter $\epsilon$.
The smallness of $\epsilon$ is understood if the shined value of $<\chi>$
is assumed on matter brane\cite{shine}.
Assuming that $<\chi>\sim M$ at the distant brane and their mass in the 
bulk is about $\sim M$, the suppression factor is given by the shining 
method:\footnote{
See ref.\cite{shine} for more detail.}
\begin{equation}
\epsilon \sim \frac{e^{-M r_{B}}}{r_{B}^{n_{E}-2}} 
\end{equation}
for $n_E >2$ and $r_B M \gg 1$, where $n_{E}$ denotes the number of the
relevant extra dimensions.
For $n_E =2$ and $r_B M \gg 1$,
\begin{equation}
\epsilon \sim \frac{e^{-M r_B}}{\sqrt{M r_B}}.
\end{equation}
Thus one can obtain the $e^{-M r_B}$ suppression for each $\epsilon$.
Here $r_{B}$ denotes the distance between the baryon number breaking
brane and the matter brane.
In this case, because of the suppression $\epsilon^{k}$, baryon number
violating A-terms are safely suppressed by the exponential factor at the
true vacuum in order not to produce dangerous operators.
On the other hand, because the suppression factor originates from the
brane distance $r_{B}$, such A-terms are not suppressed when branes are
on top of each other.\footnote{
Of course one can assume that the baryon number
breaking hidden brane is identical to the supersymmetry-breaking hidden
brane.
In such a case, the brane distance $r_{susy}$ is identified with $r_B$.}
Let us consider an example where a higher dimensional term with the 
lowest $k$ determines the phase of the Affleck-Dine condensate in the
true vacuum, while
other therms dominate when $r_B =0$.
The phases of these direct contacting A-terms are in general different
from the 
one at the true vacuum, thus producing the misalignment of the phase
during thermal brane inflation. 
Because these alternative contributions become tiny right after the end
of thermal brane inflation, misalignment of the phase is expected to
appear just after thermal brane inflation.

Of course one can expect the case where the thermal brane inflation
does not modify the baryon number violating operators in the
superpotential.
This happens when the smallness of the
operator is produced by other mechanisms that are not relevant to the brane
distance, or in the case where $r_{B}$ is not modified during thermal
inflation.
In this case the modification of the A-term is induced only by the
supersymmetry breaking, which is precisely the same as what
happens in the conventional Affleck-Dine baryogenesis.

In both cases, one can expect that the baryogenesis starts at 
$r_{B} \sim M^{-1}$ or $r_{susy} \sim M^{-1}$, where the exponential
suppression becomes 
significant.
The calculation of the resultant baryon to entropy ratio is similar to
the conventional Affleck-Dine baryogenesis.
Here we assume that the A-term of the form
\begin{equation}
V_A \simeq \frac{a m_{soft}}{M}\phi_{AD}^4
\end{equation}
is already recovered at the beginning of the AD oscillation.
Taking the initial amplitude $\phi_{AD}^{ini}\sim M$, we
obtain\cite{AD_dine} 
\begin{equation}
\frac{n_{B}}{s}\sim \frac{T_{R2}}{H_o} 
\frac{|a m_{soft} (\phi_{AD}^{i})^4|}{M\, \rho_I}\delta_{eff}.
\end{equation}
Here the inflaton density is denoted by $\rho_{I}\sim M^{4}$.
Then we can obtain  
\begin{equation}
\label{baryon0}
\frac{n_{B}}{s}\sim 10^{-10} \left(\frac{a}{10^{-7}}\right)
\left(\frac{T_{R2}}{10 GeV} \right)
\left(\frac{10^{8}GeV}{M}\right)^3.
\end{equation}

Of course in some cases the inflaton density may be determined by the
scale of the supersymmetry-breaking auxiliary component $F_X$, 
such as $\rho_I \sim |F_X|^2$.
In this case the baryon to entropy ratio becomes about 
$O\left(\left|\frac{M}{F_X^{1/2}}\right|^4\right)$ times
larger than eq.(\ref{baryon0}).

The most significant difference from the conventional Affleck-Dine
baryogenesis with extra dimensions is the absence of the problematic 
 suppression factor that makes it impossible to 
realize Affleck-Dine baryogenesis on the brane\cite{AD_extra}.

\section{Conclusions and Discussions}
\hspace*{\parindent}
In this paper we have considered an alternative mechanism of Affleck-Dine
baryogenesis that starts after thermal brane inflation.
Our mechanism works in models with supersymmetry breaking at
the distant brane.
The brane distance is required to be modified during thermal brane inflation
in order to activate the  alternative source of supersymmetry breaking. 
Besides the thermal inflation that we have considered in this paper, 
there are many models in which the initially reduced extra dimensions are
used to prevent difficulties related to the large extra
dimensions\cite{other}. 
Extensions to these models will be discussed in the next
publication\cite{next}.

\section{Acknowledgment}
We wish to thank K.Shima for encouragement, and our colleagues in
Tokyo University for their kind hospitality.

\end{document}